# South American auroral reports during the Carrington storm

Hisashi Hayakawa[1,2,3,4*], José R. Ribeiro[5], Yusuke Ebihara[6,7], Ana P. Correia[5] and Mitsuru Sôma[8]

**Abstract**

The importance of the investigation of magnetic superstorms is not limited to academic interest, because these superstorms can cause catastrophic impact on the modern civilisation due to our increasing dependency on technological infrastructure. In this context, the Carrington storm in September 1859 is considered as a benchmark of observational history owing to its magnetic disturbance and equatorial extent of the auroral oval. So far, several recent auroral reports at that time have been published but those reports are mainly derived from the Northern Hemisphere. In this study, we analyse datable auroral reports from South America and its vicinity, assess the auroral extent using philological and astrometric approaches, identify the auroral visibility at −17.3° magnetic latitude and further poleward and reconstruct the equatorial boundary of the auroral oval to be 25.1° ± 0.5° in invariant latitude. Interestingly, brighter and more colourful auroral displays were reported in the South American sector than in the Northern Hemisphere. This north–south asymmetry is presumably associated with variations of their magnetic longitude and the weaker magnetic field over South America compared to the magnetic conjugate point and the increased amount of magnetospheric electron precipitation into the upper atmosphere. These results attest that the magnitude of the Carrington storm indicates that its extent falls within the range of other superstorms, such as those that occurred in May 1921 and February 1872, in terms of the equatorial boundary of the auroral oval.

## Introduction

The Carrington storm is one of the benchmarks for space weather events that accompanied the earliest observations of a white-light flare (Carrington 1859; Hodgson 1859), and is one of the largest magnetic storms in observational history (Tsurutani et al. 2003; Cliver and Dietrich 2013; Hayakawa et al. 2019). The intensity of the white-light flare has been estimated to be ≈ X45 ± 5 in the soft X-ray (SXR) class on the basis of amplitude of its synchronised magnetic crochets at Kew and Greenwich, producing a large $\Delta H$ amplitude of ≈ −110 nT (Stewart 1861; Bartels 1937; Cliver and Dietrich 2013; Curto et al. 2016). This storm had an estimated interplanetary coronal mass ejection (ICME) transit time of ≈ 17.6 h, indicating an extremely high velocity (Cliver and Svalgaard 2004; Gopalswamy et al. 2005; Freed and Russell 2014; c.f., Knipp et al. 2018).

The ICME resulted in an extremely large magnetic storm, probably due to the combination of its high velocity, density, and strong southward interplanetary magnetic field (Gonzalez et al. 1994; Daglis et al. 1999). Magnetic disturbances with extreme deviations associated with this storm have been reported even in mid- to low-magnetic latitudes. The Colaba magnetogram revealed a negative excursion of ≈ −1600 nT in the horizontal force (hereafter $H$) and provided the basis for its intensity estimation, suggested as DST ≈ −1760 nT based on a spot value (Tsurutani et al. 2003). Its hourly Dst index was estimated as ≈ −900 (+50, −150) nT (Siscoe et al. 2006; Gonzalez et al. 2011; Cliver and Dietrich 2013). This extreme deviation has been controversially

*Correspondence: hisashi@nagoya-u.jp
[1] Institute for Space-Earth Environmental Research, Nagoya University, Nagoya 4648601, Japan
Full list of author information is available at the end of the article associated with the enhancement of the ring current (Tsurutani et al. 2003; Li et al. 2006; Keika et al. 2015), contribution of the auroral electrojet (Akasofu and Kamide 2005; Green and Boardsen 2006; Cliver and Dietrich 2013), and the field-aligned current (Cid et al. 2014, 2015).

Other mid-latitude magnetograms were also significantly affected and went off scale in the *H* component at various Russian stations and at Collegio Romano (*e.g.* Nevanlinna 2008; Blake et al. 2020). Furthermore, a large deviation of the declination (*D*) was reported, for example, in the city of Antigua, Guatemala (Ribeiro et al. 2011). Investigations and documentation on such magnetic superstorms are important not just for scientific interest, as their effects on modern civilisation could be catastrophic because of our increasing dependency on technology-based infrastructure (Baker et al. 2008; Oughton et al. 2017; Riley et al. 2018; Boteler 2019; Hapgood 2019; Zesta and Oliveira 2019; Oliveira et al. 2020).

During the Carrington superstorm, the aurorae and stable auroral red (SAR) arcs extended toward the equator. The observational reports were compiled by Kimball (1960) based on original earlier reports, such as the report by Loomis (1860); see also Shea and Smart (2006). Further investigations have revealed additional datable auroral observations from Australia (Humble 2006), Russia (Hayakawa et al. 2019), East Asia (Hayakawa et al. 2016, 2019), Mexico (González-Esparza and Cuevas-Cardona 2018), and ship logs (Green and Boardsen 2006; Hayakawa et al. 2018b). These reports were analysed on the basis of the angular distance of the observational site from the north magnetic pole (magnetic latitude; hereafter MLAT), the maximal elevation angle of the reported auroral display, and the estimated equatorial boundary of the auroral oval or auroral emission region in the Northern Hemisphere as ≈ 30.8° or 28.5° in the MLAT, at the footprint of the magnetic field line using both datable records and those without exact dates. In combination with the magnetic disturbance, the auroral equatorial boundary characterise this storm as one of the most extreme cases, rivalled only by other superstorms such as those that occurred in May 1921 and February 1872 (Silverman and Cliver 2001; Silverman 2008; Cliver and Dietrich 2013; Hayakawa et al. 2018a, 2019; Love et al. 2019).

In this context, it is still of particular interest to reconstruct the equatorial auroral boundary in the Southern Hemisphere, given that most discussions focus on auroral reports regarding the Northern Hemisphere. Among them, the datable observations in South America are of significant importance, as the reports from Santiago and Valparaíso have resulted in the most equatorial auroral observations in the Southern Hemisphere. However, in the South American sector, only a few reports from Chile have been considered in discussions on auroral visibility associated with the Carrington event, in non-Spanish literature (Kimball 1960; Wilson 2006; Hayakawa et al. 2019), apart from an undatable Columbian report (Moreno Cárdenas et al. 2016). Moreover, only Loomis (1860, pp. 398–399) and Heis (1860, pp. 37–38) translated the *El Mercurio de Valparaíso* newspaper reports as part of their compilation. Therefore, we extend the investigations involving contemporary South American literature on auroral reports in Spanish and examine their scientific implications.

## Observations

We examined newspapers and scientific bulletins regarding the Carrington storm in Chile and Argentina (see Appendix 1), as we did not find any reports from Brazil (Hayakawa et al. 2019). To date, we have identified six series of auroral reports from Chile. Two of them were translated from *El Mercurio de Valparaíso* (MV1 and MV2 in Appendix 2) and have also been incorporated in Loomis (1860, pp. 398–399) and Heis (1860, pp. 37–38), as previously indicated. Loomis's source newspaper for Concepción is not yet identified. The other records originated from the *Anales de la Universidad de Chile*, wherein one record mentioned a naval report from "the crew of the brigantine Dart that sailed around latitude S19° and longitude W149° of Greenwich" (see Appendix 3).

Our investigation is summarised in Table 1 and Fig. 1. These reports are mostly derived from observations in the most populated area of Chile, between Valparaíso (S33° 06′, W71° 37′) and Nacimiento (S37° 30′, W72° 40′). We located one naval report by Dart (S19°, W149°), a Chilean ship sailing back from Papeete, Tahiti Island, in French Polynesia (*Le Messenger*, 1859-09-04). The magnetic latitude, angular distance between the observational sites and the dipole axis, in 1859 were computed to be between − 21.8° MLAT and − 26.2° MLAT in Chile and − 17.3° MLAT for the vessel Dart, based on the GUFM1 archaeomagnetic field model (Jackson et al. 2000). Near Australia, HMS Herald witnessed aurorae around Mellish Reef in the Coral Sea (− 25.3° MLAT; Appendix 4). In contrast, we found no relevant reports in Argentina or Brazil, although the absence of evidence is no evidence of absence. Indeed, contemporary Chileans also reported this absence of data back in 1861, stating: "yet nothing was reported by the newspapers of Peru and the states of *Mar del Plata* [NB: Argentina and Uruguay]" (AUC, v. 19, p. 333; Appendix 3). To date, the naval report of the Dart has confirmed the equatorial boundary of the auroral visibility to − 17.3° MLAT, surpassing the existing reports

**Table 1 Auroral observations in South America and its vicinity**

| Year | Month | Date | Reference | Site | Latitude | Longitude | MLAT | Start | End | Direction | Colour |
|---|---|---|---|---|---|---|---|---|---|---|---|
| 1859 | 9 | 1 | L4-14 | Concepción | S36° 46′ | W73° 02′ | − 25.5 | Midnight | 26 | | R |
| 1859 | 9 | 1 | L4-15 | Santiago | S33° 28′ | W70° 40′ | − 22.1 | 26 | 29 | s | B/R/Y |
| 1859 | 9 | 1 | L4-15 | Valparaíso | S33° 06′ | W71° 37′ | − 21.8 | | | | |
| 1859 | 9 | 1 | AUC1 | Santiago | S33° 28′ | W70° 40′ | − 22.1 | 25 | 29 | s | Pi/B/Y |
| 1859 | 9 | 1 | AUC1 | Concepción | S36° 46′ | W73° 02′ | − 25.5 | 25 | 29 | s | Pi/B/Y |
| 1859 | 9 | 1 | AUC1 | Nacimiento | S37° 30′ | W72°40′ | − 26.2 | | | | |
| 1859 | 9 | 1 | AUC2 | Santiago | S33° 28′ | W70° 40′ | − 22.1 | 25 | 29 | s-z-nw | R/W/Pi/B |
| 1859 | 9 | 1 | AUC2 | Dart | S19° | W149° | − 17.3 | | | | |
| 1859 | 9 | 1 | AUC3 | Santiago | S33° 28′ | W70° 40′ | − 22.1 | 25 | 28 | s-w | Pi/R/W |
| 1859 | 9 | 1 | AUC3 | Yumbel | S36° 46′ | W72° 34′ | − 25.5 | 24 | 27 | | R |
| 1859 | 9 | 1 | AUC4 | Rancagua | S34° 10′ | W70° 45′ | − 22.8 | | | | |
| 1859 | 9 | 1 | NYHW | Levant | S50° | W78° | − 38.8 | 25 | | s-all | R |
| 1859 | 9 | 2 | Herald | Herald | S17° | E156° | − 25.3 | 19 | 25 | ese-sws | R |

Their date, reference (see Appendix 1), observational site with their latitude and longitude, magnetic latitude, start and end of their visibility in local mean time (LMT), direction, and colour are given. The directions are shown in n (north), s (south), e (east), w (west), and all (all sky). The colourations are described as R (red), Pi (pink), W (white), B (blue including purple), and Y (yellow). Their LMT is shown in 6 -- 30 (6 on the following day; i.e., LMT+24), to describe the observational time beyond midnight continuously from Honolulu at 20.5° MLAT, naval observations at 22.8° MLAT, and Valparaíso at − 21.8° MLAT.

In Chile, the auroral display was first reported over Concepción near midnight and by other cities between 1.5 h and 2 h on 2 September in LMT. The display reportedly lasted until 4 h LMT or even later and was obfuscated during twilight, which was slightly earlier than the calculated onset of astronomical twilight at ≈ 4.9 LMT. Its colour was reported as primarily reddish, pinkish, and bluish or purplish, with occasional mentions of yellowish or whitish regions, whereas it was compared to fire and flame in *El Mercurio de Valparaíso*. These colour patterns confirm that these displays were not a form of stable auroral red (SAR) arcs (Hayakawa et al. 2018a; see also Kozyra et al. 1997). The reported auroral visibility in the South American sector in 24–29 LMT is converted to 19.5–25.5 MLT. This visibility duration is located in the dusk sector to the midnight sector. The reported aurorae are not associated with discrete aurorae nor diffuse aurorae (Lui et al. 1973; Akasofu 1974) but with low-latitude aurorae (Shiokawa et al. 2005), because of the dominance of the reddish coloration. Their extremely low MLAT (25–29) hinders us from applying analogy of high-latitude aurorae.

Two reports described the auroral extension at Santiago (S33° 28′, W70° 40′; − 22.1° MLAT). Wenceslao Diaz reported that the auroral display at approximately 1.5–2 h LMT "invaded almost all the Southern Hemisphere of the sky and a great portion of the northern region" (AUC, v.19, p. 331). At approximately 2.5 h LMT, the aurora developed to its maximum: "Over this gloomy part rose an immense luminous arch: its ends coincided with those of the above mentioned dark band and its circumference disappeared to the East in the *Argo Navis* constellation, to the North in the *Eridanus*, and to the West in the constellations of *Grus*, *Sagittarius*, *Aquila*, *Lyra* and *Sagitta*" (AUC, v.19, p. 331; Appendix 3). Another witness in Carlos Huidobro's report commented: "At about 2:00 a.m., it rose to its maximum height, covering about one-third of the celestial dome, above the meridian of Santiago, and reaching to the western horizon of this part of the sky" (AUC, v.19, p. 340; Appendix 3).

## Equatorial auroral boundary

Diaz and Huidobro described the auroral extent slightly differently. Diaz stated that the aurora covered "almost all the Southern Hemisphere of the sky" and stretched beyond the zenith, whereas Huidobro described the maximum auroral height as "covering about one-third of the celestial dome, above the meridian of Santiago, and reaching to the western horizon of this part of the sky." Based on their descriptions, the auroral elevation was determined as ≥ 90° and ≈ 60°, respectively. This apparent discrepancy should be considered further.

Interestingly, Diaz outlined the details of the auroral extent in comparison with the constellations when the event reached its maximum strength (≈ 2.5 LMT). Accordingly, the circumference of the immense luminous arch over the main auroral emission "disappeared to the East in the Argo Navis constellation, to the North in the Eridanus, and to the West in the constellations of Grus, Sagittarius, Aquila, Lyra and Sagitta" and a "purple

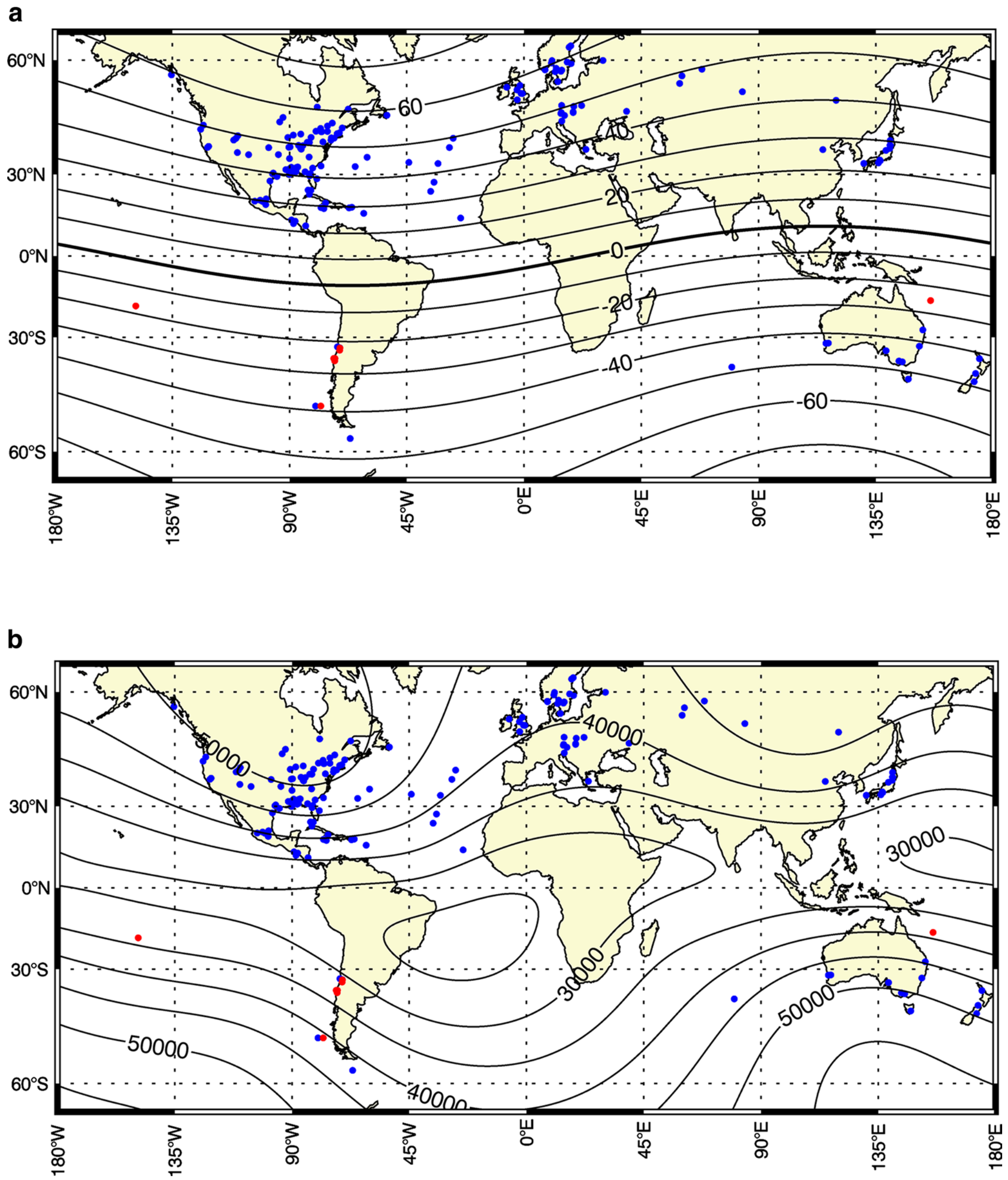

**Fig. 1** Auroral visibility on 1/2 September 1859, reported in this article (red dots), in comparison with the known auroral reports (blue dots) (see Hayakawa et al. 2019). The contour lines in the top panel indicate the magnetic latitude, and those in the bottom panel indicate the intensity of the magnetic field at 400 km altitude, which was computed using the GUFM1 model (Jackson et al. 2000)

transparent gauze" covered Centaurus, Crucis, Canopus, and the Magellan Clouds (AUC, v.19, p. 331).

Simulation of the star positions reveals their estimated locations in Fig. 2. Among them, Lyra (Lyr) and Sagitta (Sge) were below the horizon and are omitted from the discussion. Aquila (Aql) is slightly in the northern sky near the western horizon. Assuming that the auroral height was 220–400 km (see *e.g.* Roach et al. 1960; Ebihara et al. 2017) and using the geomagnetic pole determined by the GUFM1 model (Jackson et al. 2000), we plotted the dipole field lines shown in Fig. 3, which could represent the auroral display that Diaz described. The western and northern borders are constrained by the lower edge formed by Aql and Eri. This constraint is satisfied with the magnetic field lines spanning $L = 1.21–1.30$ (24.6°–28.7° for the invariant latitude (ILAT; see O'Brien et al. 1962; Fig. 2 of Hayakawa

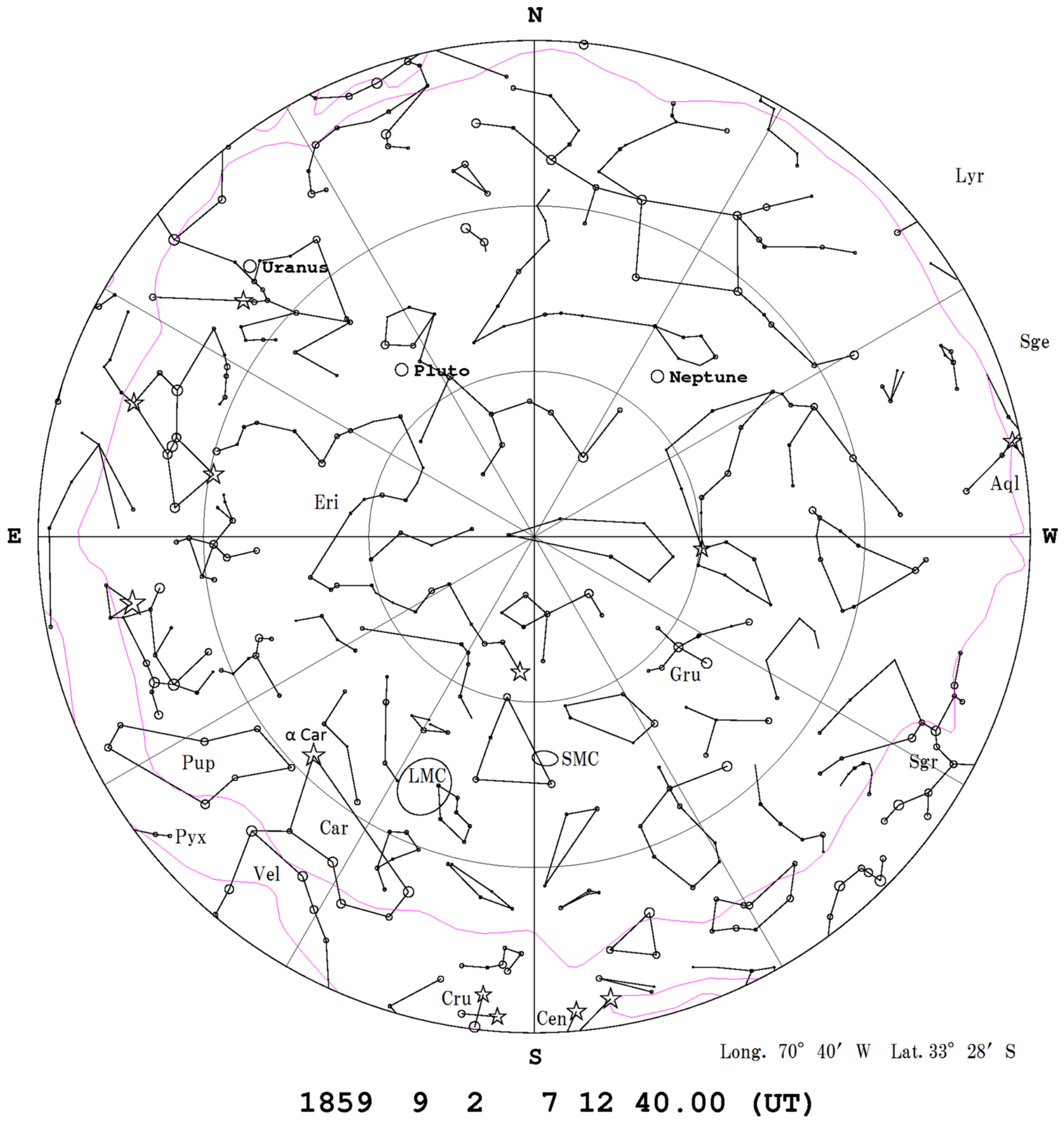

**Fig. 2** Simulated star positions at 2.5 h LMT on 2 September 1859, as viewed from Santiago (S33° 28′, W70° 40′). This is shown using an all-sky view with its zenith at the centre and its geographic north at the top. The constellations are abbreviated as Eri (Eridanus), Car, Vel, Pup (Argo Navis), Gru (Grus), Sgr (Sagittarius), Aql (Aquila), Lyr (Lyra), Sge (Sagitta), Cen (Centaurus), Cru (Crucis), *α* Car (Canopus), SMC (the Small Magellan Cloud), and LMC (the Large Magellan Cloud). The two constellations of Sge and Lyr are outside the map limits, as they were below the horizon. The constellations are abbreviated as Eri (Eridanus), Gru (Grus), Sgr (Sagittarius), Aql (Aquila), Lyr (Lyra), Sge (Sagitta), Cen (Centaurus), and Cru (Crux). The Argo Navis constellation is now divided into the 3 constellations Car (Carina), Vel (Vela), and Pup (Puppis). Canopus is the brightest star in Car and it is denoted by α Car. The Large and Small Magellan Clouds are denoted by LMC and SMC, respectively

et al. 2018a)) Another possible scenario that sets its western edge at the western horizon suggests a slightly more conservative reconstruction with field lines spanning $L = 1.23–1.30$ (25.6°–28.7° ILAT). Therefore, these reports imply that the equatorial auroral boundary in the South American sector lies between 24.6° and 25.6° ILAT. Hereinafter, we refer to it as $25.1 \pm 0.5°$ ILAT for simplicity.

## Contextualisation of the Chilean reports

The equatorial auroral boundary ($25.1° \pm 0.5°$ ILAT) is more equatorward than those estimated from observations made in the Northern Hemisphere (30.8° ILAT or 28.5° ILAT) on the basis of the datable Sabine report or the Honolulu report without an exact date (Hayakawa et al. 2018a) with the same altitude assumption of ≈ 400 km (Roach et al. 1960; Ebihara et al. 2017). Regarding the north–south asymmetry, there are two possible reasons. First, observations in both hemispheres

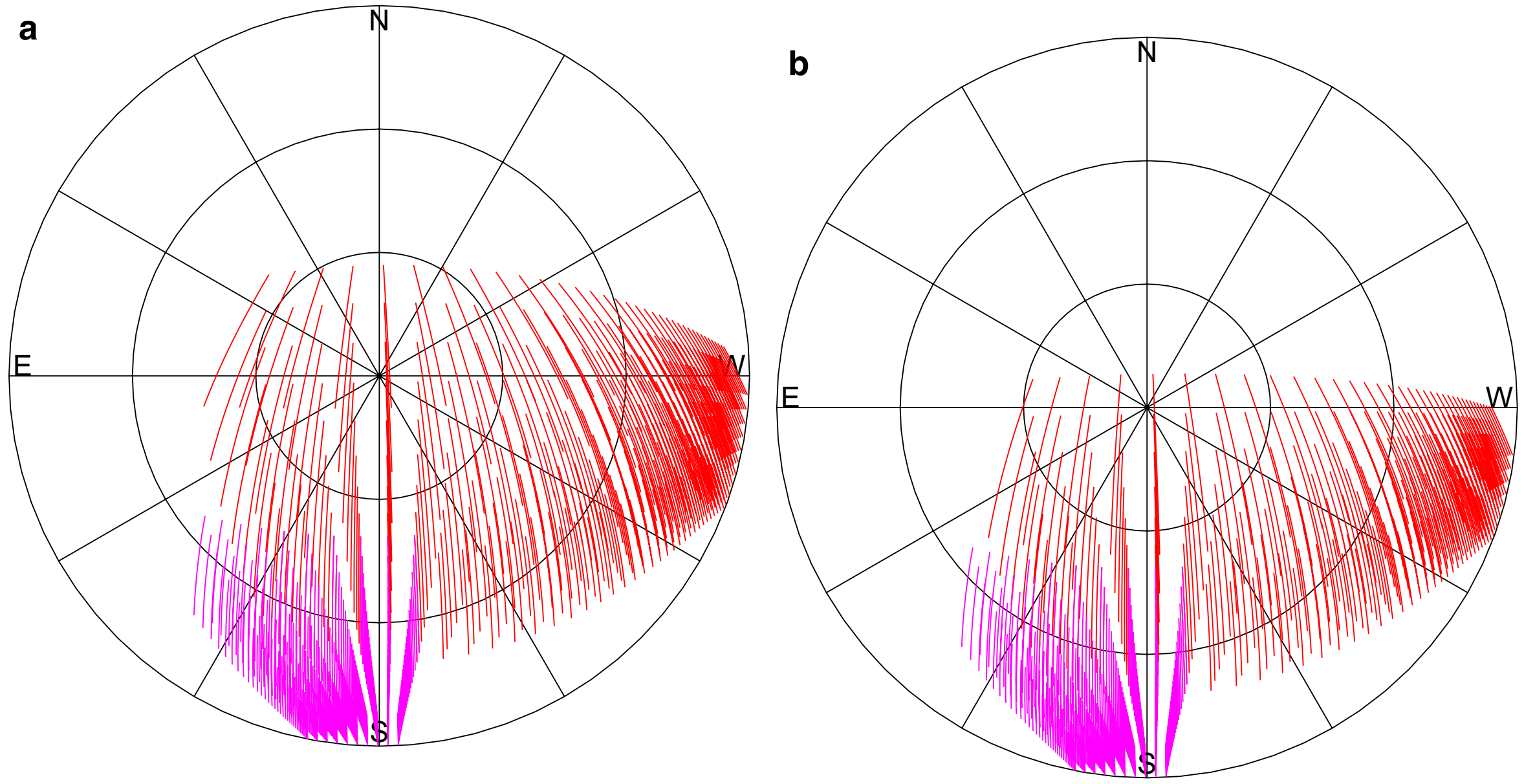

**Fig. 3** Dipole field lines that could represent the auroral display observed at S33° 28′, W70° 41′. The red lines, which could manifest the reddish aurora, are the field lines ranging between $L = 1.21$ and 1.30 (24.6° and 28.7°) in the left panel, and those ranging between $L = 1.23$–1.30 (25.6° and 28.7° ILAT) with heights ranging from 220 km to 400 km. They are constrained by the constellations observed at 2.5 LMT on 2 September 1859 (Diaz's report) or by the auroral edge at the western horizon (Huidobro's report). In both cases, these red lines span from − 20.0° to − 2.5° in the magnetic longitude (MLON). The purple lines (that could manifest the purple transparent gauze") range between $L = 1.29$ and 1.60 (28.3° and 37.8° ILAT) with the same height as the red lines, span from − 5.0° to − 0.5°

were not made simultaneously. This may be partially because of the difference of the MLON; Santiago at − 4.4° MLON (25.5–29 LMT = 6.2–9.7 GMT), Sabine at − 19.1° MLON (24.5–27 LMT = 6.1–8.6 GMT), and Honolulu at − 96.4° MLON (22 LMT–, with dating uncertainty = 8.5 GMT–). While the auroral visibilities at Santiago and Sabine were reported almost simultaneously, their different MLONs (14.7° MLON) equivalent to ≈ 1 MLT and the elliptic shape of the auroral oval can produce a few degrees of displacement in the equatorial auroral boundary.

This is more typically the case with the variation of the equatorial auroral boundaries in South America and Australia. Near the Australian sector, HMS Herald reported auroral visibility up to 25° in elevation (see Appendix 4). This allows us to estimate the equatorial boundary of the auroral oval as 34.5° ILAT based on the assumption of the said altitude threshold. This is most probably because of its MLON difference and local night-time corresponding to the storm recovery phase, in comparison with the Santiago report with its local night corresponding to the storm main phase (see also Fig. 3 of Hayakawa et al. 2019). This is probably the case with the East Asian sector in a similar MLON, where aurorae were similarly reported in the storm recovery phase (Figure 8 of Hayakawa et al. 2016; Figure 7 of Hayakawa et al. 2019). Nevertheless, absence of their exact elevation angles still hinders us from reconstructions of the equatorial auroral boundary in this sector.

Second, the magnitude of the magnetic field was different and may have differed the amount of precipitating electrons (*e.g.* Stenbaek-Nielsen et al. 1973). According to the GUFM1 model with an epoch of 1859, the magnitude of the magnetic field was 30782 nT at 400 km above Santiago (see Fig. 1b), whereas the magnitude was 34035 nT at 400 km above the magnetic conjugate point (N06° 54′, W065° 48′). The magnitude in the Southern Hemisphere was approximately 90% of that in the Northern Hemisphere. The magnitude of the magnetic field at the apex is 17,525 nT. Considering the electrons located at the apex of the magnetic field line, the loss-cone angles ($\alpha_L$) of the northbound and southbound electrons were calculated to be 45.9° and 49.0°, respectively. For electrons with an isotropic pitch angle distribution, the solid angle of the loss cone ($= 2\pi$ (1-cos $\alpha_L$)) is 1.91 str for the northbound electrons, whereas it is 2.16 str for the southbound electrons. This implies that trapped electrons can precipitate lower into the upper atmosphere over Santiago than at the conjugate point, because of the relatively weak mirror force. Consequently, it is speculated that a greater number of electrons precipitated into the Southern Hemisphere, resulting in brighter and more colourful aurorae over Santiago than at the conjugate point. This speculation is based on the assumption that electrons with energy > 1 keV selectively precipitate into the upper atmosphere in the Southern Hemisphere. Ebihara et al. (2017) showed the precipitating electron flux having two peaks at 71 eV and 3 keV during the extreme events of March 1989 and October 2003. Further studies are required to confirm the speculation on the basis of statistical studies of the precipitating electrons.

As such, our reconstruction of the equatorial auroral boundary (25.1° ± 0.5° ILAT) in the Southern Hemisphere shifts more equatorward than previously reported (30.8° or 28.5° ILAT; Hayakawa et al. 2018b). This is comparable to the superstorms that occurred in February 1872 and May 1921. On 4 February 1872, the aurora was reported up to the zenith of Shanghai (19.9° MLAT), and hence, the equatorial boundary of its auroral oval was estimated to be ≈ 24.2° ILAT. In contrast, the Colaba magnetogram indicates a magnetic disturbance of Dst* ≤ − 830 nT (Hayakawa et al. 2018a). On 14/15 May 1921, the magnetic superstorm caused an extreme disturbance of Dst* ≈ − 907 ± 132 nT (Love et al. 2019), and an aurora was reported up to an altitude of 22° by the Apia Observatory in Samoa (Angenheister and Westland 1921, p. 202; Silverman and Cliver 2001). Given that the MLAT by Apia was estimated to be − 16.2°, the equatorial boundary of the auroral oval was estimated to be ≈ 27.1° ILAT (Hayakawa et al. 2019). The equatorial boundary of the auroral oval during these superstorms, the Carrington storm (≈ 25.1° ± 0.5° ILAT) the May 1921 storm (≈ 27.1° ILAT), and the February 1872 storm (≈ 24.2° ILAT) agree with existing comparisons related to the intensity of their magnetic disturbances (Cliver and Dietrich 2013; Hayakawa et al. 2019; Love et al. 2019).

## Conclusion

In this article, we analysed datable auroral reports from South America during the Carrington storm. Our analyses provided further details of the auroral displays observed in Chilean cities and nearby vessels. These reports provided more data on the low-latitude aurorae observed in the Southern Hemisphere and updated the equatorial boundary of the auroral oval during this storm, based on naval observations by the vessel Dart at − 17.3° MLAT. The multiple colourations mentioned in the Chilean reports suggest that they were not SAR arcs, but were auroral emissions instead, despite their low MLATs. Moreover, Diaz and Huidobro reported on the extent of the auroral display observed over Santiago (− 22.1° MLAT). In contrast, Huidobro described the aurora as "covering about one-third of the celestial dome", and hence they probably extended over as much over the sky as ≈ 60°. In contrast, Diaz provided the auroral extent in comparison with the constellations. Based on his descriptions, we computed their positions and estimated the auroral extent as ≥ 90°. Accordingly, the equatorial boundary of the auroral oval in the Southern Hemisphere at that time was reconstructed as ≈ 25.1 ± 0.5° ILAT.

This is relatively more equatorward than the boundary reconstructions in the Northern Hemisphere of 30.8° or 28.5° ILAT, on the basis of datable reports with/without the Honolulu report without clear dating (Hayakawa et al. 2018b). The north–south asymmetry has been associated with the difference between the observational time and the magnitude of the magnetic field. While this was almost simultaneous with the low-latitude observations in the Caribbean Sea such as Sabine, these observations had difference of MLT ≈ 1. The calculation using GUFM1 also indicates that the magnitude of the magnetic field at Santiago in 1859 was almost 10% lower than at the conjugate point. We suppose that trapped electrons precipitated into the upper atmosphere over Santiago in greater numbers than over the conjugate point in the Northern Hemisphere because of the relatively weak mirror force. These differences probably caused the brighter and more colourful auroral displays observed over Santiago compared to those observed in the Northern Hemisphere.

Our results indicate that the equatorial boundary of the auroral oval during the Carrington storm is comparable to that of the February 1872 storm and the May 1921 storm boundaries, and agrees with the results based on the intensity of the hourly Dst* estimates. Such extension of the auroral oval represents a threat to modern civilisation, as the approach the auroral oval enhances magnetic disturbances and triggers geomagnetically induced currents (Boteler et al. 1998, 2019; Pulkkinen et al. 2012; Riley et al. 2018; Hapgood 2019). Indeed, Secchi has described an extreme deviation of ≈ 3000 nT for the Roman magnetogram, which was presumably under or near the auroral oval (Cliver and Dietrich 2013; Blake et al. 2020). The equatorial boundary of the auroral oval of 25.1 ± 0.5° ILAT covers almost all major cities, not only in Europe and the United States, but also in the northern half of East Asia. Given our increasing dependence on technology based on electronic infrastructure, the consequences of such extreme storms could be disastrous, particularly because the Carrington storm is not an isolated storm in the observational history, but likely only one of the several possible magnetic superstorms, as evidenced by the two subsequent storms that occurred in 1872 and 1921 (*e.g.* Cliver and Dietrich, 2013; Hayakawa et al. 2019; Love et al. 2019; Chapman et al. 2020).

**Abbreviations**
ICME: Interplanetary coronal mass ejection; ILAT: Invariant latitude; LMT: Local mean time; MLT: Magnetic local time; MLAT: Magnetic latitude; SAR arc: Stable auroral red arc.

**Acknowledgements**
We thank the Biblioteca Pública Santiago Severin, Valparaíso (Nelson Cortés Rojas), the Service do Patrimoine Archivistique et Audiovisuel de la Polynésie, Papeete, and the National Archives of the United Kingdom at Kew for providing copies of the historical documents, and the Biblioteca Pública General San Martín, Mendoza (Rolando Landabour) and the Biblioteca Pública Universidad de la Plata (Mario Carnabali, Patricia Lischinsky y María Marta Isla) for the search through Argentine newspapers. We thank Ana I. Ribeiro for her helpful comments. The authors benefited from discussions within the ISSI International Team #510 (SEESUP Solar Extreme Events: Setting Up a Paradigm) and ISWAT-COSPAR S1-02 team.

**Authors' contributions**
HH designed and constructed this manuscript. JRR and APC collected and translated the South American reports. YE supervised this study and conducted simulations. MS conducted astrometrical calculations. All authors read and approved the final manuscript.

**Funding**
This study has been financially supported by JSPS Grant-in-Aids JP15H05812, JP20H00173, JP20K20918, and JP17J06954, JSPS Overseas Challenge Program for Young Researchers, the 2020 YLC collaborating research fund, and the research grants for Mission Research on Sustainable Humanosphere from Research Institute for Sustainable Humanosphere (RISH) of Kyoto University, Young Leader Cultivation (YLC) program of Nagoya University, the Unit of Synergetic Studies for Space of Kyoto University, and the BroadBand Tower.

**Availability of data and materials**
References and translations of historical documents are shown in Appendices.

**Competing interests**
Not applicable.

**Author details**
[1] Institute for Space-Earth Environmental Research, Nagoya University, Nagoya 4648601, Japan. [2] Institute for Advanced Research, Nagoya University, Nagoya 4648601, Japan. [3] UK Solar System Data Centre, Space Physics and Operations Division, RAL Space, Science and Technology Facilities Council, Rutherford Appleton Laboratory, Harwell Oxford, Didcot, Oxfordshire OX11 0QX, UK. [4] Nishina Centre, Riken, Wako 3510198, Japan. [5] Escola Secundária Henrique Medina, Esposende Av. Dr. Henrique Barros Lima, 4740-203 Esposende, Portugal. [6] Research Institute for Sustainable Humanosphere, Kyoto University, Uji 6100011, Japan. [7] Unit of Synergetic Studies for Space, Kyoto University, Kyoto 6068306, Japan. [8] National Astronomical Observatory of Japan, Mitaka 1818588, Japan.

## Appendices

### Appendix 1: Bibliography of observational reports

| | |
|---|---|
| AUC1 | *Anales de la Universidad de Chile*, v. 16, Santiago, Universidad de Chile, 1859 |
| AUC2 | *Anales de la Universidad de Chile*, v. 18, Santiago, Universidad de Chile, 1861 |
| AUC3 | *Anales de la Universidad de Chile*, v. 19, Santiago, Universidad de Chile, 1861 |
| Herald | *Journal of Herald's proceedings in the Coral Sea &c, H.M. Denham (ff.1–188)*, ADM 55/73, the National Archives of the United Kingdom at Kew |
| MV1 | *El Mercurio de Valparaíso*, 1859-09-02, p. 3 |
| MV2 | *El Mercurio de Valparaíso*, 1859-09-05, p. 2 |
| NYHW | *NY Harper's Weekly*, 1859-12-10, p. 787 |

### Appendix 2: English translations of *El Mercurio de Valparaíso* (MV1 and MV2)

MV1: *El Mercurio de Valparaíso*, 1859-09-02, p. 3

There are people who say they saw in Valparaíso the same phenomenon that was observed in Santiago on Thursday night. They say the atmosphere was covered by a red cloak like fire.

MV2: *El Mercurio de Valparaíso*, 1859-09-05, p. 2

Extraordinary phenomenon – It is really interesting the atmospheric phenomenon that was observed yesterday at dawn in Santiago. In this regard, today's *Ferrocarril* says:

"Yesterday's morning, about 1:30 or 2:00, the atmosphere to the south of Santiago was seen extraordinarily illuminated with a bright light, coloured pink, blue and yellow. This strange appearance, as we have said before, kept much of our population alarmed and even shocked, because it is entirely unknown and its cause cannot be explained. Some of the people who witnessed it believed at first that it was a great fire in San Bernardo, whose glow managed to illuminate much of the atmosphere. This phenomenon remained visible for about three hours.

### Appendix 3: English translations of the AUC reports

#### *AUC1*

METEOROLOGY–Apparition of an aurora australis in Santiago and Concepción.–by Ramón Briseño

At about 1:30 or 2:00 in the morning of the 2nd of September, we saw the atmosphere, towards the south of both cities, extraordinarily illuminated by a pink, blue, and yellow light, with the shape of a cloud or a globe of marsh fire, which released some kind of flame or vapour and spread out with a clarity similar to that of the Moon, and whose movement was contrary to that of the Earth. This strange meteorological phenomenon, which remained visible for about three hours, did not fail to alarm the population, no doubt for being almost totally unknown in this area. Meanwhile, almost certainly, it was what is called the aurora borealis in the Northern Hemisphere, and in our Hemisphere, we call the aurora australis. (…) (a)

(a) Later, we were informed that the same phenomenon was also seen in Nacimiento.

***AUC2***

METEOROLOGY–Data on the aurora polaris in both hemispheres on the night of 2nd September 1859, communicated to the Faculty of Physical and Mathematical Sciences. (pp. 328–359)

I. Aurora australis observed in Santiago on the morning of 2nd of September 1859 by Mr. Wenceslao Diaz (pp. 330–340)

(…)

pp. 331–332

After moonset, the night stayed in its normal darkness. Then, between 0:30 and 1:00 a.m., it began to appear to the southwest and above the horizon, a red light quite similar to what results in certain circumstances from the decomposition of sunlight, which half an hour later invaded almost all the southern hemisphere of the sky and a great part of the northern hemisphere.

By 2:30 a.m., the phenomenon achieved its maximum development, together with the greatest show of light. Above the hills to the south and to the west of this city, a wide dark band stretched from the SW to the NW, seemingly made of the haze that sometimes arises above the horizon in the coldest nights. The central part of this band, covering ¼ of its length, formed the base of a dark circular segment that seemed of the same nature and whose centre was situated more or less in the direction of the southwest. Over this gloomy part rose an immense luminous arch: its ends coincided with those of the above-mentioned dark band and its circumference disappeared to the east in the *Argo Navis* constellation, to the north in the *Eridanus*, and to the west in the constellations of *Grus*, *Sagittarius*, *Aquila*, *Lyra,* and *Sagitta*. The colour of the luminous arch, in the part that crowned the dark segment, was of intense carmine, which softened through insensitive gradations until it became red in its central part, becoming a beautiful light pink colour spilling into reddish and whitish shades through the celestial dome. Through this purple transparent gauze, the stars acquired a golden colour, and among them, the brightest were *Centaurus* and *Crucis*, Canopus, a few of second magnitude, and the Magellan Clouds. The glow of this phenomenon illuminated the atmosphere with a diffuse light, and the roofs of the buildings were stained as if with the last rays of the evening twilight.

The deepest silence reigned; the sky was completely clear and serene; the stars shone in all their splendour; there was not the slightest gust of wind, and the atmosphere was mild on this November night.

The culminating part of the dark segment was approximately 15° above the horizon, to the west of the magnetic meridian, while that of the luminous arch was below *Eridanus*.

At 2:25 a.m., the pink colour slowly intensified, but in five minutes it returned in the same way to its first state; shortly after, the same change was repeated, and these alternating changes in colour continued, seeming to have their origin in the dark segment, from where they radiated to the circumference until 4:00 a.m., when the light dimmed and extinguished, obfuscated by the dawn of the new day.

During the persistence of the phenomenon, I observed a declination needle that remained fixed and without any signs of oscillation; I marked its position in order to see later if during the night it had suffered some deviation; but, it continued to point out the same position until 12:00 a.m., when I left this observation.

(…)

p. 332

Identical observations (*similar changes in the colours of the aurora*) were made by the crew of the brigantine Dart that sailed around latitude 19°S and longitude 149°W of Greenwich.

(…)

p. 333

Great has been its (*the aurora*) extension, as can be deduced from what I have observed; and although we lack the observations that may have been made in the Magellan Strait and in Australia, we must assume their presence in these regions, taking into account the austral situation they occupy. The observers of the Dart observed it all night in the low latitude of 19°S, and yet nothing appeared in the newspapers of Peru or the states of Mar del Plata (*N.B. Argentina and Uruguay*).

(…)

II. Other data on the same aurora australis in Chile (pp. 340–341)

Data were collected in Santiago by Mr. Carlos Huidobro, from witnesses who observed the aurora: Mr. Domingo Tagle, Mr. Nazario Salas, Mr. Moises del Fierro, and Mr. Fernando de la Plata.

The aurora appeared around 1:00 a.m., beginning to appear towards the southeast part of the horizon, a very light pink light, which was gaining more and more height, and changing in colour from pink to a blood red. Through this light, all the stars could be seen. At about 2:00 a.m., it rose to its maximum height, covering about one-third of the celestial dome above the meridian of Santiago, reaching to the western horizon of this part of the sky. It then remained steady, retaining its fire-red colour, more shadowy in the middle part, and diminishing in intensity on the east and west sides, in such a way that there was nothing left on the horizon but a light pink hue, the same as the hue when it first appeared.

At 3:30 a.m., all this light began to decline towards the west, diminishing slowly in intensity until dawn, after which it took on a whitish colour that was lost in the light of day.

Data collected by Mr. Aníbal Pinto, reported by a trustworthy person who observed this aurora on the plain five leagues north of Yumbel (Latitude of Concepción).

From the point where this person was located, one could see a luminous arch, one of whose stars rested in Antuco (*N.B. A volcano to the southeast*). The arch ran from east to west along the south horizon. It was formed of a bright red band of a non-dark red, limited in its lower part by a black ribbon approximately 1/30 of the width of the previous ribbon. The intensity of the light of this strip was similar to the moon, although somewhat sheltered. It was observed from 0:00 until after 3:00 a.m. It is assured by the witness that, before seeing the aurora, this part of the sky was dark.

### *AUC3*

METEOROLOGY–Atmospheric phenomenon occurred in Santiago de Chile the 26th of July, 1861–Communication from priest Enrique Cappelletti, S. J., to the Faculties of Physical Sciences and Medicine in the session of the 10th of September of the same year.

p. 344

(…) when happened here a few years ago, a similar phenomenon (*Enrique Cappelletti uses the resemblance of the phenomenon he saw with the 1859 event to argue that what he saw now was also an aurora*) was recognized as an aurora australis, and was seen from Rancagua, which is to the south of Santiago, and people thought it was a fire, and from Santiago it was seen towards the magnetic pole. (…)

## Appendix 4: Transcription of HMS Herald's report (pp. 214–216)

On one of these nights (September 2nd) of extreme anxiety under evolutions to "old ones own" in such doubtful waters. –We were beguiled into extra watchfulness by, to us in this region, the unusual appearance of the sky which at one hour after sunset until an hour after midnight presented a ruby tinted field of thin gauzy clouds as if in front of the azure sky, the foot of which was lifted 4° degrees above the horizon while the head of the curtain was elevated 25° degrees; its expanse was from ESE to SSW; all its margins were fleecy, it displayed no varied tints–no golden hues as if of zodiacal order–but as if reflecting a vast conflagration or some active Volcano? Of the later the nearest I knew was that of "Tanna" of the New Hebrides distant 750 miles in line with the left part of the phenomenon from us.

It was a serenely clear night with the moon in her first quarter right opposite descending from an altitude. The uniform ruby tint dissipated rather suddenly at 1 am when the sky became generally charged with threatening clouds which induced reefing our canvas; but neither rain nor boisterous weather ensued throughout the following day. The Barometer which stood at 30.002 at 6 pm rose to 30.048 at midnight and then fell to 30.000 at 1.45 a.m. The temperature of air and sea (75° of Fahrenheit) continued the same throughout.

## Publisher's Note